\documentclass{vgtc}                          

\graphicspath{{figures/}{pictures/}{images/}{./}} 

\usepackage{times}                     

\DeclareUnicodeCharacter{0097}{-}

\usepackage{tabu}                      
\usepackage{booktabs}                  
\usepackage{lipsum}                    
\usepackage{mwe}                       

\usepackage{mathptmx}                  

\onlineid{0}

\vgtccategory{Research}

\vgtcinsertpkg

\title{Playing telephone with generative models: ``verification disability,'' ``compelled reliance,'' and accessibility in data visualization}

\author{Frank Elavsky\thanks{e-mail: fje@cmu.edu}\\ %
        \scriptsize Carnegie Mellon University %
\and Cindy Xiong Bearfield\thanks{e-mail: cxiong@gatech.edu}\\ %
     \scriptsize Georgia Tech %
     }

\teaser{
  \centering
  \includegraphics[width=\linewidth, alt={A source image visualization is shown: Money lost each year to theft in billions. Then chatGPT to chatGPT shows 3 visualizations in succession. Claude to Claude also shows 3 visualizations in succession. These visualizations are unpacked in greater detail in the appendices.}]{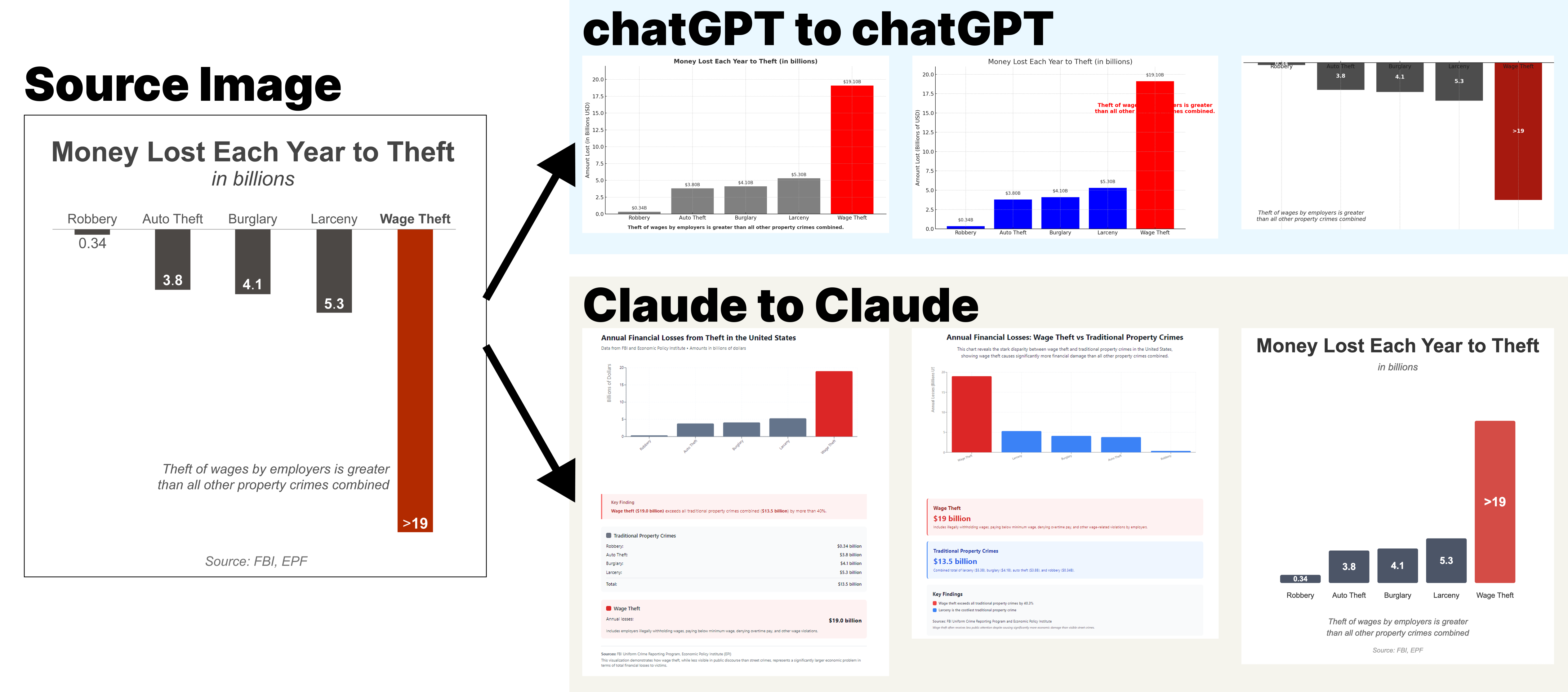}
  \caption{In a game of telephone, we had models describe a visualization and then a new model instance had to create a visualization based only on the first model's description. We recursively repeated this 3 times across 2 different models. Models produced a variety of different kinds of bias in the descriptions and subsequent image constructions, some with specifically troublesome ramifications for blind users in particular.}
  \label{fig:teaser}
}

\abstract{
    This paper is a collaborative piece between two worlds of expertise in the field of data visualization: accessibility and bias. In particular, the rise of generative models playing a role in accessibility is a worrying trend for data visualization. These models are increasingly used to help author visualizations as well as generate descriptions of existing visualizations for people who are blind, low vision, or use assistive technologies such as screen readers. Sighted human-to-human bias has already been established as an area of concern for theory, research, and design in data visualization. But what happens when someone is unable to verify the model output or adequately interrogate algorithmic bias, such as a context where a blind person asks a model to describe a chart for them? In such scenarios, trust from the user is not earned, rather reliance is \textit{compelled} by the model-to-human relationship. In this work, we explored the dangers of AI-generated descriptions for accessibility, playing a game of telephone between models, observing bias production in model interpretation, and re-interpretation of a data visualization. We unpack ways that model failure in visualization is especially problematic for users with visual impairments, and suggest directions forward for three distinct readers of this piece: technologists who build model-assisted interfaces for end users, users with disabilities leveraging models for their own purposes, and researchers concerned with bias, accessibility, or visualization.
} 


\begin{document}


\firstsection{Introduction}

\maketitle

The visualization community has long recognized the importance of addressing human-to-human bias in data representation and interpretation \cite{xiong2019curse}. Yet as the visualization community increasingly delegates the work of visualization interpretation to generative AI systems, we risk creating new forms of bias that are more subtle, more systematic, and more difficult to detect and correct~\cite{mehrabi2021survey, chatzimparmpas2025visual}.

The rapid integration of generative AI models into accessibility tools presents a profound shift in how people with disabilities interact with data visualizations~\cite{seo2024maidr, gorniak2024vizability}. While these technologies promise to democratize access to visual information~\cite{oswal2024examining}, they simultaneously introduce new forms of bias that threaten to undermine the very accessibility they claim to enhance~\cite{tang2025everyday}. This intersection of generative AI, accessibility, and data visualization demands urgent attention from our research community, not merely as a technical challenge to be solved, but as a fundamental threat to equitable access to information.

A doctor reviewing patient mortality rates, a policymaker analyzing crime data, and a financial advisor studying market trends can all depend on data visualizations to make critical decisions. 
But when a blind or low-vision user encounters the same visualizations, they may have no choice but to rely on a generative model's interpretation. 
Suddenly, an AI system becomes the sole mediator between life-altering information and high-stake decisions.

Unlike traditional accessibility tools that provide structured, verifiable access to data, or when descriptions are human-authored and author-verified, generative models operate as black boxes, producing descriptions that users cannot independently validate. This creates a dangerous power asymmetry: the model's interpretation becomes the user's reality, regardless of its accuracy, completeness, or embedded biases. 
We refer to this process with asymmetrical power dynamics as
``compelled reliance,'' where users are provided no reasonable alternatives except to rely on the model output.


The current landscape of AI-assisted accessibility tools suffers from a troubling lack of accountability mechanisms~\cite{novelli2024accountability}. When a model generates an incorrect or biased description, who bears responsibility? The model developer? The application creator? The institution deploying the technology? This diffusion of responsibility creates an accountability vacuum that leaves users with disabilities particularly vulnerable to harm with no path forward for justice~\cite{cooper2022accountability}.

To demonstrate the danger of compelled reliance on visualization and chart descriptions created by generative models, we played a game of `telephone' with two popular LLM models: ChatGPT (model 4o) and Claude (Sonnet 4)~\cite{openai2023chatgpt, anthropic2024claude}. 
Telephone is a game where information is passed from one person to the next, and at the end it loops back to the originator of the information. 
We demonstrate how information lost or manipulated upstream can produce unsolvable problems downstream in the context of visualization design and chart description generation.

In our experimentation, we started with a human-generated visualization, then asked for a description, sent the description to a model to create a visualization, used the visualization as a prompt to ask for a new description, and so on. We didn't experiment with models intending to \textit{prove} that bias exists (other work has already explored generative model bias, such as~\cite{hong2025llms, wang2024aligned, bendeck2024empirical}), but rather to expose observations-as-provocations for discussion.
We offer this demonstration as a call to action for AI, accessibility, and visualization researchers.
We aim to catalyze much-needed conversations about the ethical implications of using generative AI for visual analytics and push our community toward more responsible development, deployment, interrogation, and even \textit{non-use} of these powerful technologies.



\section{Why does machine bias in visualization accessibility matter?}
People often rely on heuristics and biases to make decisions~\cite{kahneman2011thinking}, but access to visualizations and visual analytics promises more evidence-based insights and potentially more optimal data-driven decisions.

However, even in the most pristine data analytics pipeline, bias is inevitably woven into every stage of authoring visualizations, whether the work is conducted by humans or machines~\cite{wall2017warning, gaba2023my, ren2024bias, zhao2018bias, dimara2018task}.
For example, the collection and omission of data is a well-documented source of bias. D'Ignazio and Klein's work on `capta' demonstrates how what gets counted and what doesn't can shape our understanding of social phenomena~\cite{d20204}. 
In visual analytics, missing data patterns can mislead viewers when not properly encoded~\cite{wall2017warning, song2018s}. 
The sifting, sorting, and cleaning of data can further introduce additional biases~\cite{zgraggen2018investigating}.
When the data is ready to be visualized, design choices, such as selecting a chart type, a color palette, or a textual annotation, can prime viewers toward specific conclusions~\cite{bearfield2023does, liu2018somewhere, li2025confirmation}.
Design artifacts to enhance data accessibility, such as generating captions or alt-texts for blind and low-vision users, often well-intentioned (and even necessary for access), can nonetheless emphasize different aspects of the same data, potentially exacerbating algorithmic biases and social biases~\cite{lundgard2021accessible}.

It's all bias, all the way down.

So ``are we doomed?'' as Michael Correll has remarked multiple past occasions~\cite{correll2025personal}, or does this present an opportunity for the visualization community to take action against the expanding constellation of biases that generative AI has introduced to human-data interaction?






\subsection{People with disabilities are already marginalized}

The material harms and risks of bias are often, unfortunately, assumed by whoever was led astray by the artifact or person who biased them in the first place. And in the case of machine-generated descriptions of data visualizations for accessibility (and any case of machine-automation in accessible design), people with disabilities will ultimately assume the risks and harms of model failure~\cite{MacLeod2017, Whittaker2019, Bennett2020, Bennett2021}.

This matters to highlight because people with disabilities are already disproportionately marginalized by a growing technological divide~\cite{Whittaker2019}, suffer from the existing lack of access to vital systems that enable participation in life~\cite{Bennett2020}, must learn and adapt in more complex ways to emerging technologies~\cite{Lundgard2019,Ymous2020,Makati2024}, develop complex coping strategies to deal with technological accessibility barriers~\cite{Andrew2025, Makati2024}, and yet are often considered ``costly'' and ``burdensome'' by latent and explicit ableism in socio-technical politics~\cite{Lundgard2019,Spiel2020,Sum2024}.

In the context of visual analytics, having inaccurate descriptions of data can risk even greater material harm for people with disabilities~\cite{MacLeod2017, Bennett2021}.
We argue that greater care should be taken to interrogate model failure and put more interventions and guardrails in place for accessibility design artifacts of data visualizations.

More work should be done, and more care taken. As a matter of equity and justice, people with disabilities should be spared from as much harm from bias as possible. 

\subsection{Responsibility and accountability is diffused}

Presently, if a human authors a chart description that leads to harm, exclusion, waste, or some other cost, we can hold that person (or their employer) accountable. Someone can be sued when people harm the lives of people with disabilities.

But when models are used, responsibility becomes diffused~\cite{Widder2023}. People are less likely to verify the information that a model produces, and in cases where systems are fully automated, no human can be held accountable~\cite{AttardFrost2025,Widder2023}. We conjecture that if human authors don't believe they are responsible for the harms caused by bias they produce, they will not invest as much care in ensuring the accuracy and truthfulness of their work.

Who should be held accountable for machine-produced bias in visualization accessibility? Of course, we argue that tool and model authors are partly responsible, and so are the tool and model users. But if responsibility doesn't neatly and transparently map to decisions made in the production of a data visualization and its accessibility artifacts (such as a description), then justice (through accountability) becomes diffused and obscured.

In an ideal world, authors take responsibility for doing their work well, minimizing harm wherever possible. When harm does occur, there is transparency and clarity about who is accountable for restoration and remediation, especially when justice is pursued.

\section{Model-to-model telephone}
We didn't intend to run an \textit{experiment} per se, but rather conduct \textit{experimentation} sessions of model-based interaction where we intentionally elicit examples of model failure. We wanted to create an environment where bias production upstream produces downstream epistemological concerns.

Telephone is a game played by passing a message through whispering in one person's ear. That person then passes the message to another, and so on. This process repeats for a number of people (all players in the game) and eventually the message returns to the original person. The game is a good illustration of how even one person incorrectly hearing or relaying information early on inherently produces unverifiable bias in all other players that follow them in the game.

We believe that this game is a good environment for observing ways that models produce the same kind of information collapse, because agents in a game of telephone are inherently subjected to \textit{verification disability} and \textit{compelled reliance} on whoever passes that agent information.

\subsection{``Verification disability'' and ``compelled reliance''}
Our theoretical scenario of model failure and bias production are contexts where humans are presented information but are ``blind'' (figuratively or literally) to the meaningful artifacts worth interrogating.

If humans are not provided appropriate methods and means of applying skepticism and calibrating trust of information that is given by a model, then that model actually produces disability. ``Disability'' in this sense is defined by the \textit{social model of disability}, where environmental factors, not a person's body, produce barriers in a person's ability to do something, such as perform a task, participate, or otherwise act on their own agency~\cite{Shakespeare2006, Barnes2019}.

So all of the problems we observed in the following section and subsections are a result of contexts where models produce \textit{verification disability}: they do not provide adequate means for someone to be able to test, validate, and calibrate trust for the information given. In some contexts, the user might be literally blind and unable to interrogate the accuracy of a description of a visual artifact. In other contexts, especially ones that are based on humanless workflows (model to model or fully-automated flows, in particular), then the system also produces barriers to a human's ability to verify the information produced by that system. Our work here should then be interpreted as relevant not only for contexts where an end user is blind, but any context where a person is not provided an ability to adequately verify information produced by a generative model.

So in cases where adequate means of verification are not accessible, models then produce an environment of \textit{compelled reliance}. We define ``compelled reliance'' as a context where users have no reasonable alternative but to entirely trust the results of a model. This problem arises in cases where models are mostly correct. Consistent distrust of a mostly-correct system produces worse overall outcomes than consistent trust, so it is generally more reasonable for a user to trust a model than not. \textit{Compelled reliance} then becomes dangerous in the cases when a model \textit{is} biased but, due to lack of verifiability, cannot be held accountable or responsible for negative outcomes for anyone who inherits its biases.

\subsection{Experimentation set-up}
Our set-up was simple: we chose a data visualization produced by our first author in 2017, which was popular on Reddit at the time \cite{reddit2017}. The data visualization was an assembly of different data sources taken from the Economic Policy Institute (EPI) and the Federal Bureau of Investigations (FBI) on the topic of wage theft, comparing projected data on wage theft to reported cases of other kinds of theft (robbery, auto theft, burglary, and larceny).

Notable features of the original visualization (see ``Source Image'' in  \Cref{fig:teaser}):
\begin{enumerate}
    \itemsep0em 
    \item The bars grow down, instead of up, to semantically communicate "loss" of money instead of positive quantities.
    \item The visualization was designed using Adobe Illustrator, so it is relatively polished compared to most charting tools.
    \item The visualization has ``$>$19'' for the data on the bar for Wage Theft, because this was the lowest bound in the model used to extrapolate by the EPI. Exact values are unknown.
    \item The visualization has an annotation near the wage theft bar that explains that "theft of wages by employers is greater than all other property crimes combined."
    \item The visualization has a typo, which cites the sources as "FBI, EPF" even though the correct sources are "FBI, EPI."
\end{enumerate}

We fed this graphic into a model (both chatGPT's gpt-4o \cite{openai2023chatgpt} and Claude's Sonnet 4 \cite{anthropic2024claude}) with the prompt: ``can you describe this?'' 

Next, we took the resulting description and started a new chat, asking the model ``Can you make a chart based on this description?''

We repeated these two steps another time, each using a new chat instance. Each instance of chat was ensured to be free from knowledge of previous chats (using user settings to ensure previous chats were unconnected).

Our final step was to ask the model at the end, ``This description does not accurately represent the image. What is missing? Why did this happen?'' and, after its response ask, ``Can you try to recreate the original, fixing our visualization using the original’s design choices?''

Each flow then involved 2 generated descriptions, 2 generated visualizations, 1 generated description that was intended to be ``corrected,'' 1 final ``corrected'' visualization, and then a further discussion (with Claude) about who should be responsible for model failure in this context.

We had 4 different flows for our game of telephone: chatGPT and chatGPT, Claude and Claude, Claude for descriptions and chatGPT for images, and chatGPT for descriptions and Claude for images.

For the sake of brevity in this short piece, we will be skipping the results and discussion of our cross-model experimentation. The chatGPT-chatGPT and Claude-Claude results give us plenty to discuss, which you can find a record of in our appendices (see \Cref{sec:appendix-a} and \Cref{sec:appendix-b}). Additionally, the important characteristics of model failure when using a cross-model approach were not notably different from our same-model approach.

\section{Observing Generative Model Failures}
The point of this paper is not to provide a systematic analysis of model failure within a controlled experiment. Instead, our work is intended to explore an emerging and urgent problem, in hopes that our peers will become inspired enough to follow up and address these problems rigorously. For this reason, our observations of model failures should be interpreted as preliminary and provocative.

We will explain categorically different kinds of bias we observed, as well as why we believe each kind of bias is particularly harmful. All of the observations can be found in our appendices (see \Cref{sec:appendix-a}, \Cref{sec:appendix-b}, and  \Cref{sec:appendix-c}).
    
\subsection{Information loss}

In our examples, the models lost information when building a graphic based on a description. The $>$ (greater than) symbol and its meaning was not transferred to subsequent graphics and the data source was lost (no longer an annotation or referenced in the image). chatGPT added 0.1 to 19 (making it 19.1) while claude eventually lost all precision (we surmise because it did not have direct labels on bars, subsequent descriptions were less accurate). Claude's final dataset was 5.5, 4.5, 4, 0.5, and 19, when the correct values were 5.3, 4.1, 3.8, 0.34, and $>$19.

\subsection{Information fabrication}

Interestingly, the typo in the original image for ``EPF'' (should have been ``EPI'') was corrected without mention by both models immediately. This raises concerns about cases when a model will produce incorrect tokens or when a model remediates information without notifying the user. It is possible that the original data really did come from a source labeled ``EPF'' or that knowing there was a typo was an important detail to mention.

In our later conversations with Claude (see \Cref{sec:appendix-c}), Claude considered the data sources EPI and FBI as ``confident'' when questioned, but did not provide explanations. However, in our last conversation with chatGPT and Claude (see the end of \Cref{sec:appendix-a} and \Cref{sec:appendix-b}, respectively), both models actually fabricated the data source as ``Economic Policy Foundation'' when asked to make a description specifically for someone who is blind.

Additionally, chatGPT produced a nearly-accurate description of the otherwise unreadable annotation added to its second image, which was surprising. We surmise it was able to do this because chatGPT's setting to avoid chat history is either false (and it actually shares history between chats, despite appearing not to) or the specific phrase ``theft of wages from workers is greater than all other forms of theft crimes combined'' was a strong enough association to come through, despite missing many words.

\subsection{Incorrect emphasis and interpretation of purpose}

As the original author of the original image, my motivation for creating the graphic was to stimulate discussion on the lack of data that exists on wage theft. We don't track it as part of the FBI's Uniform Crime Reporting and as a result, our best guess of loss comes from projections by the Economic Policy Institute's study.

While it is unfair to assume a model would be able to accurately state \textit{why} this graphic was created, both models nevertheless still stated reasons for why they thought this chart was impactful, important, and so on. By not authoring a description myself (as the author), the emphasis of purpose was left up to the model.

\subsection{Poor design comprehension}

The original image broke common visualization convention, showing a downward direction to the bars. Both models originally did not recognize and did not replicate it. Additionally, chatGPT argued that the ``color and size'' of the wage theft bar was used to emphasize its importance. But the size was determined by the data value, it was not a design decision like color. 

Additionally, in the phase of our process where we questioned the models and asked them to recreate the original more faithfully, chatGPT placed labels in an undreadable fashion while Claude stated that it would replicate a ``stacked/building-block visual approach - bars appear to build upward rather than starting from a common baseline.'' Of course, the bars are not stacked or ``building-blocks'' and in fact build \textit{down} while starting at a common baseline.

While not an artifact of our telephone game, chatGPT's bar chart places gridlines over the bars by default, which is not an ideal design choice. Overall, text hierarchy was quite poorly handled until Claude's final image attempt.


\subsection{Under-description}

Interestingly, when asked to describe the visualization specifically for someone who was blind, both models omitted significant information from the description. When questioned, the models clarified that it was ``best practice'' to keep alt text short, despite not asking for ``alt-text'' but instead simply a description of the chart for someone who is blind. (``Alt-text'' is a specific technical artifact that provides a way for authors to describe images in HTML, which isn't the same as an image description, broadly-speaking.)

\subsection{Over-confidence}

Of course to no surprise, the models were quite confident in their assessments and in the graphics they produced. This over-confidence is dangerous, because the original data source from the EPI report was not. By default, neither model offered any degree of transparency about the uncertainty of their descriptions and images generated.

When later questioned, Claude interestingly decided to clarify what it was confident about and what it wasn't confident about, but the items that it mentioned it wasn't confident about were all previously discussed in the prompting process with a high degree of certainty.

\section{A call for future work}
This position paper is intended as a provocation to our peers. If human-to-model and model-to-model interactions run the risk of bias production, then new research and innovation should engage ways for users to validate information and hold model creators accountable, not just prompt further. Alternatives should be explored that do not solely rely on generative models as a fix for the problems that generative models produce.

\vspace{2mm}
\noindent \textbf{Tool creators and tool users}
Tool creators and tool users should bear responsibility for their work~\cite{Widder2023}. 
To improve the quality of life for people with disabilities, practitioners (and companies) that build visualization tools (in addition to authors that leverage those tools), must put care into producing excellent, truthful, and meaningful data visualizations, holding their work to a high standard.

Practitioners should also consider forgoing the use of generative model assistance in accessibility entirely at times~\cite{Zong2024}, especially if there is an opportunity to gain more design expertise and consideration for people with disabilities, rather than delegating this work. By forgoing the use of a model, authors gain the opportunity to learn how to describe images themselves and also gain the opportunity to provide information that a model cannot: context about why the visualization mattered to create in the first place~\cite{Elavsky2022Tools, lundgard2021accessible}.

Additionally, companies should have support in place for practitioners and users with disabilities to seek justice and support when systems fail. And ample transparency about risks involved in model use should be communicated to tool-users and end-users.

\vspace{2mm}
\noindent \textbf{Users with disabilities}
Users with disabilities should remain skeptical of the information provided by models when parsing and creating data visualizations. Models alone do not provide enough resources for validating the contents of a visualization. Some older, slower methods, such as access to a data table or rendered data~\cite{bartram2021untidy, Zong2022, Elavsky2023}, sonification~\cite{walker2010universal,Thompson2023, Kim2024}, or the traditional method of asking a friend or colleague for assistance~\cite{Bennett2018,Hofmann2020}, may all be useful strategies to employ in different contexts.

And accessibility data is not as portable as pixel formats, such as pngs. So descriptions may be constantly authored and re-authored as charts and graphs wander to different corners of the internet. So whenever possible and feasible, seek out the original author-provided descriptions of a data visualization. The original authors are the most responsible for telling the truth and avoiding harm.

\vspace{2mm}
\noindent \textbf{Researchers}
There is much we still don't know. Researchers must restlessly interrogate models and their failures. We also need to have more techniques studied that are not just for builders and fellow researchers, but end-users. Researchers must devote the production of knowledge at this intersection towards informing the people who are most at risk of harmful outcomes. And most importantly, partnerships with people with disabilities should be formed, so that people with disabilities can be involved in innovation and investigation of data-related accessibility~\cite{Lundgard2019,Spiel2020,Elavsky2022Chartability}.

Researchers have a unique opportunity to interrogate the lack of transparency in modern AI ecosystems as well. A large part of the ethical considerations mentioned in this piece are a result of accountability-diffusion because it is unclear who is at fault for producing harm and bias~\cite{Widder2023}. 
Through thorough investigation and well-founded recommendations for more just technical systems and policies, we can help bring clarity to those seeking justice.

\acknowledgments{
    Special thanks to Eytan Adar for a conversation (circa 2019?) about telephone and information loss. Thanks also to conversations with the AXLE lab (CMU) over the years as well, on similar topics. The authors wish to acknowledge how important conversations can be for sparking something, even years later.
}

\bibliographystyle{abbrv-doi}

\bibliography{paper}

\begin{thebibliography}{10}

\bibitem{Andrew2025}
S.~Andrew and G.~Tigwell.
\newblock Understanding the experiences of people with and without vision impairments when using mobile user interface alternative color modes.
\newblock {\em Proceedings of the ACM on Human-Computer Interaction}, 9, 09 2025. doi: {{%
10\hspace{.1pt}\discretionary{.}{%
}{.}\hspace{.4pt}1145\discretionary{/}{%
}{/}3743704}}


\bibitem{anthropic2024claude}
{Anthropic}.
\newblock Claude, 2024.
\newblock Large language model.

\bibitem{AttardFrost2025}
B.~Attard-Frost and D.~G. Widder.
\newblock The ethics of ai value chains.
\newblock {\em Big Data and Society}, 12(2), May 2025. doi: {{%
10\hspace{.1pt}\discretionary{.}{%
}{.}\hspace{.4pt}1177\discretionary{/}{%
}{/}20539517251340603}}


\bibitem{Barnes2019}
C.~Barnes.
\newblock {\em Understanding the social model of disability}, p. 14–31.
\newblock Routledge, Oct. 2019. doi: {{%
10\hspace{.1pt}\discretionary{.}{%
}{.}\hspace{.4pt}4324\discretionary{/}{%
}{/}9780429430817\discretionary{%
}{-}{-}2}}


\bibitem{bartram2021untidy}
L.~Bartram, M.~Correll, and M.~Tory.
\newblock Untidy data: The unreasonable effectiveness of tables.
\newblock {\em IEEE Transactions on Visualization and Computer Graphics}, 28(1):686--696, 2021.

\bibitem{bearfield2023does}
C.~X. Bearfield, C.~Stokes, A.~Lovett, and S.~Franconeri.
\newblock What does the chart say? grouping cues guide viewer comparisons and conclusions in bar charts.
\newblock {\em IEEE Transactions on Visualization and Computer Graphics}, 30(8):5097--5110, 2024. doi: {{%
10\hspace{.1pt}\discretionary{.}{%
}{.}\hspace{.4pt}1109\discretionary{/}{%
}{/}TVCG\hspace{.1pt}\discretionary{.}{%
}{.}\hspace{.4pt}2023\hspace{.1pt}\discretionary{.}{%
}{.}\hspace{.4pt}3289292}}


\bibitem{bendeck2024empirical}
A.~Bendeck and J.~Stasko.
\newblock An empirical evaluation of the gpt-4 multimodal language model on visualization literacy tasks.
\newblock {\em IEEE VIS}, pp. 1--11, 2024.
\newblock to appear. doi: {{%
10\hspace{.1pt}\discretionary{.}{%
}{.}\hspace{.4pt}1109\discretionary{/}{%
}{/}TVCG\hspace{.1pt}\discretionary{.}{%
}{.}\hspace{.4pt}2024\hspace{.1pt}\discretionary{.}{%
}{.}\hspace{.4pt}3456155}}


\bibitem{Bennett2018}
C.~L. Bennett, E.~Brady, and S.~M. Branham.
\newblock Interdependence as a frame for assistive technology research and design.
\newblock In {\em Proceedings of the 20th International ACM SIGACCESS Conference on Computers and Accessibility}, ASSETS ’18, p. 161–173. ACM, Oct. 2018. doi: {{%
10\hspace{.1pt}\discretionary{.}{%
}{.}\hspace{.4pt}1145\discretionary{/}{%
}{/}3234695\hspace{.1pt}\discretionary{.}{%
}{.}\hspace{.4pt}3236348}}


\bibitem{Bennett2021}
C.~L. Bennett, C.~Gleason, M.~K. Scheuerman, J.~P. Bigham, A.~Guo, and A.~To.
\newblock “it’s complicated”: Negotiating accessibility and (mis)representation in image descriptions of race, gender, and disability.
\newblock In {\em Proceedings of the 2021 CHI Conference on Human Factors in Computing Systems}, CHI ’21, p. 1–19. ACM, May 2021. doi: {{%
10\hspace{.1pt}\discretionary{.}{%
}{.}\hspace{.4pt}1145\discretionary{/}{%
}{/}3411764\hspace{.1pt}\discretionary{.}{%
}{.}\hspace{.4pt}3445498}}


\bibitem{Bennett2020}
C.~L. Bennett and O.~Keyes.
\newblock What is the point of fairness?: disability, ai and the complexity of justice.
\newblock {\em ACM SIGACCESS Accessibility and Computing}, (125):1–1, Mar. 2020. doi: {{%
10\hspace{.1pt}\discretionary{.}{%
}{.}\hspace{.4pt}1145\discretionary{/}{%
}{/}3386296\hspace{.1pt}\discretionary{.}{%
}{.}\hspace{.4pt}3386301}}


\bibitem{chatzimparmpas2025visual}
A.~Chatzimparmpas.
\newblock Visual analytics for explainable and trustworthy artificial intelligence.
\newblock {\em IEEE Computer Graphics and Applications}, 45(2):100--111, 2025.

\bibitem{cooper2022accountability}
A.~F. Cooper, E.~Moss, B.~Laufer, and H.~Nissenbaum.
\newblock Accountability in an algorithmic society: relationality, responsibility, and robustness in machine learning.
\newblock In {\em Proceedings of the 2022 ACM conference on fairness, accountability, and transparency}, pp. 864--876, 2022.

\bibitem{correll2025personal}
M.~Correll.
\newblock Personal communication, July 2022.
\newblock Personal communication.

\bibitem{d20204}
C.~D'Ignazio and L.~Klein.
\newblock 4.“what gets counted counts”.
\newblock {\em Data Feminism}, 2020.

\bibitem{dimara2018task}
E.~Dimara, S.~Franconeri, C.~Plaisant, A.~Bezerianos, and P.~Dragicevic.
\newblock A task-based taxonomy of cognitive biases for information visualization.
\newblock {\em IEEE transactions on visualization and computer graphics}, 26(2):1413--1432, 2018.

\bibitem{reddit2017}
F.~Elavsky.
\newblock {M}oney lost each year to theft [oc].
\newblock \url{https://www.reddit.com/r/dataisbeautiful/comments/88b0be/first_post_money_lost_each_year_to_theft_oc/}, 2017.
\newblock [Accessed 27-07-2025].

\bibitem{Elavsky2022Tools}
F.~Elavsky.
\newblock Picking the right tools for the job: Learning and building for data visualization and accessibility.
\newblock https://www.urban.org/research/publication/do-no-harm-guide-centering-accessibility-data-visualization, 2022.
\newblock [Accessed 01-04-2025].

\bibitem{Elavsky2022Chartability}
F.~Elavsky, C.~Bennett, and D.~Moritz.
\newblock How accessible is my visualization? evaluating visualization accessibility with chartability.
\newblock {\em Computer Graphics Forum}, 41(3):57--70, June 2022. doi: {{%
10\hspace{.1pt}\discretionary{.}{%
}{.}\hspace{.4pt}1111\discretionary{/}{%
}{/}cgf\hspace{.1pt}\discretionary{.}{%
}{.}\hspace{.4pt}14522}}


\bibitem{Elavsky2023}
F.~Elavsky, L.~Nadolskis, and D.~Moritz.
\newblock Data navigator: An accessibility-centered data navigation toolkit.
\newblock {\em IEEE Transactions on Visualization and Computer Graphics}, p. 1–11, 2023. doi: {{%
10\hspace{.1pt}\discretionary{.}{%
}{.}\hspace{.4pt}1109\discretionary{/}{%
}{/}tvcg\hspace{.1pt}\discretionary{.}{%
}{.}\hspace{.4pt}2023\hspace{.1pt}\discretionary{.}{%
}{.}\hspace{.4pt}3327393}}


\bibitem{gaba2023my}
A.~Gaba, Z.~Kaufman, J.~Cheung, M.~Shvakel, K.~W. Hall, Y.~Brun, and C.~X. Bearfield.
\newblock My model is unfair, do people even care? visual design affects trust and perceived bias in machine learning.
\newblock {\em IEEE transactions on visualization and computer graphics}, 30(1):327--337, 2023.

\bibitem{gorniak2024vizability}
J.~Gorniak, Y.~Kim, D.~Wei, and N.~W. Kim.
\newblock Vizability: Enhancing chart accessibility with llm-based conversational interaction.
\newblock In {\em Proceedings of the 37th Annual ACM Symposium on User Interface Software and Technology}, pp. 1--19, 2024.

\bibitem{Hofmann2020}
M.~Hofmann, D.~Kasnitz, J.~Mankoff, and C.~L. Bennett.
\newblock Living disability theory: Reflections on access, research, and design.
\newblock In {\em Proceedings of the 22nd International ACM SIGACCESS Conference on Computers and Accessibility}, ASSETS ’20, p. 1–13. ACM, Oct. 2020. doi: {{%
10\hspace{.1pt}\discretionary{.}{%
}{.}\hspace{.4pt}1145\discretionary{/}{%
}{/}3373625\hspace{.1pt}\discretionary{.}{%
}{.}\hspace{.4pt}3416996}}


\bibitem{hong2025llms}
J.~Hong, C.~Seto, A.~Fan, and R.~Maciejewski.
\newblock Do llms have visualization literacy? an evaluation on modified visualizations to test generalization in data interpretation.
\newblock {\em IEEE Transactions on Visualization and Computer Graphics}, 2025.

\bibitem{kahneman2011thinking}
D.~Kahneman.
\newblock {\em Thinking, fast and slow}.
\newblock Macmillan, 2011.

\bibitem{Kim2024}
H.~Kim, Y.-S. Kim, and J.~Hullman.
\newblock Erie: A declarative grammar for data sonification.
\newblock In {\em Proceedings of the CHI Conference on Human Factors in Computing Systems}, CHI ’24, p. 1–19. ACM, May 2024. doi: {{%
10\hspace{.1pt}\discretionary{.}{%
}{.}\hspace{.4pt}1145\discretionary{/}{%
}{/}3613904\hspace{.1pt}\discretionary{.}{%
}{.}\hspace{.4pt}3642442}}


\bibitem{li2025confirmation}
S.~Li, T.~J. Davidson, C.~Xiong~Bearfield, and E.~Wall.
\newblock Confirmation bias: The double-edged sword of data facts in visual data communication.
\newblock In {\em Proceedings of the 2025 CHI Conference on Human Factors in Computing Systems}, pp. 1--16, 2025.

\bibitem{liu2018somewhere}
Y.~Liu and J.~Heer.
\newblock Somewhere over the rainbow: An empirical assessment of quantitative colormaps.
\newblock In {\em Proceedings of the 2018 CHI conference on human factors in computing systems}, pp. 1--12, 2018.

\bibitem{Lundgard2019}
A.~Lundgard, C.~Lee, and A.~Satyanarayan.
\newblock Sociotechnical considerations for accessible visualization design.
\newblock In {\em 2019 IEEE Visualization Conference (VIS)}, p. 16–20. IEEE, Oct. 2019. doi: {{%
10\hspace{.1pt}\discretionary{.}{%
}{.}\hspace{.4pt}1109\discretionary{/}{%
}{/}visual\hspace{.1pt}\discretionary{.}{%
}{.}\hspace{.4pt}2019\hspace{.1pt}\discretionary{.}{%
}{.}\hspace{.4pt}8933762}}


\bibitem{lundgard2021accessible}
A.~Lundgard and A.~Satyanarayan.
\newblock Accessible visualization via natural language descriptions: A four-level model of semantic content.
\newblock {\em IEEE Transactions on Visualization and Computer Graphics}, 28(1):1073--1083, 2021.

\bibitem{MacLeod2017}
H.~MacLeod, C.~L. Bennett, M.~R. Morris, and E.~Cutrell.
\newblock Understanding blind people’s experiences with computer-generated captions of social media images.
\newblock In {\em Proceedings of the 2017 CHI Conference on Human Factors in Computing Systems}, CHI ’17, p. 5988–5999. ACM, May 2017. doi: {{%
10\hspace{.1pt}\discretionary{.}{%
}{.}\hspace{.4pt}1145\discretionary{/}{%
}{/}3025453\hspace{.1pt}\discretionary{.}{%
}{.}\hspace{.4pt}3025814}}


\bibitem{Makati2024}
T.~Makati, G.~W. Tigwell, and K.~Shinohara.
\newblock The promise and pitfalls of web accessibility overlays for blind and low vision users.
\newblock In {\em The 26th International ACM SIGACCESS Conference on Computers and Accessibility}, ASSETS ’24, p. 1–12. ACM, Oct. 2024. doi: {{%
10\hspace{.1pt}\discretionary{.}{%
}{.}\hspace{.4pt}1145\discretionary{/}{%
}{/}3663548\hspace{.1pt}\discretionary{.}{%
}{.}\hspace{.4pt}3675650}}


\bibitem{mehrabi2021survey}
N.~Mehrabi, F.~Morstatter, N.~Saxena, K.~Lerman, and A.~Galstyan.
\newblock A survey on bias and fairness in machine learning.
\newblock {\em ACM computing surveys (CSUR)}, 54(6):1--35, 2021.

\bibitem{novelli2024accountability}
C.~Novelli, M.~Taddeo, and L.~Floridi.
\newblock Accountability in artificial intelligence: What it is and how it works.
\newblock {\em Ai \& Society}, 39(4):1871--1882, 2024.

\bibitem{openai2023chatgpt}
{OpenAI}.
\newblock Chatgpt, 2023.
\newblock Large language model.

\bibitem{oswal2024examining}
S.~K. Oswal and H.~K. Oswal.
\newblock Examining the accessibility of generative ai website builder tools for blind and low vision users: 21 best practices for designers and developers.
\newblock In {\em 2024 IEEE International Professional Communication Conference (ProComm)}, pp. 121--128. IEEE, 2024.

\bibitem{ren2024bias}
Y.~Ren, S.~Guo, L.~Qiu, B.~Wang, and D.~J. Sutherland.
\newblock Bias amplification in language model evolution: An iterated learning perspective.
\newblock {\em Advances in Neural Information Processing Systems}, 37:38629--38664, 2024.

\bibitem{seo2024maidr}
J.~Seo, S.~S. Kamath, A.~Zeidieh, S.~Venkatesh, and S.~McCurry.
\newblock Maidr meets ai: Exploring multimodal llm-based data visualization interpretation by and with blind and low-vision users.
\newblock In {\em Proceedings of the 26th International ACM SIGACCESS Conference on Computers and Accessibility}, pp. 1--31, 2024.

\bibitem{Shakespeare2006}
T.~Shakespeare.
\newblock {\em The Social Model of Disability}, p. 16–24.
\newblock Routledge, Mar. 2006. doi: {{%
10\hspace{.1pt}\discretionary{.}{%
}{.}\hspace{.4pt}4324\discretionary{/}{%
}{/}9781003082583\discretionary{%
}{-}{-}3}}


\bibitem{song2018s}
H.~Song and D.~A. Szafir.
\newblock Where's my data? evaluating visualizations with missing data.
\newblock {\em IEEE transactions on visualization and computer graphics}, 25(1):914--924, 2018.

\bibitem{Spiel2020}
K.~Spiel, K.~Gerling, C.~L. Bennett, E.~Brulé, R.~M. Williams, J.~Rode, and J.~Mankoff.
\newblock Nothing about us without us: Investigating the role of critical disability studies in hci.
\newblock In {\em Extended Abstracts of the 2020 CHI Conference on Human Factors in Computing Systems}, CHI ’20, p. 1–8. ACM, Apr. 2020. doi: {{%
10\hspace{.1pt}\discretionary{.}{%
}{.}\hspace{.4pt}1145\discretionary{/}{%
}{/}3334480\hspace{.1pt}\discretionary{.}{%
}{.}\hspace{.4pt}3375150}}


\bibitem{Sum2024}
C.~M. Sum, F.~Spektor, R.~Alharbi, L.~B. Baltaxe-Admony, E.~Devine, H.~A. Dixon, J.~Duval, T.~Eagle, F.~Elavsky, K.~Fernandes, L.~S. Guedes, S.~Hillman, V.~Kameswaran, L.~Kirabo, T.~Motahar, K.~E. Ringland, A.~Schaadhardt, L.~Scheepmaker, and A.~Williamson.
\newblock Challenging ableism: A critical turn toward disability justice in hci.
\newblock {\em XRDS: Crossroads, The ACM Magazine for Students}, 30(4):50–55, June 2024. doi: {{%
10\hspace{.1pt}\discretionary{.}{%
}{.}\hspace{.4pt}1145\discretionary{/}{%
}{/}3665602}}


\bibitem{tang2025everyday}
X.~Tang, A.~Abdolrahmani, D.~Gergle, and A.~M. Piper.
\newblock Everyday uncertainty: How blind people use genai tools for information access.
\newblock In {\em Proceedings of the 2025 CHI Conference on Human Factors in Computing Systems}, pp. 1--17, 2025.

\bibitem{Thompson2023}
J.~R. Thompson, J.~J. Martinez, A.~Sarikaya, E.~Cutrell, and B.~Lee.
\newblock Chart reader: Accessible visualization experiences designed with screen reader users.
\newblock In {\em Proceedings of the 2023 CHI Conference on Human Factors in Computing Systems}, CHI ’23, p. 1–18. ACM, Apr. 2023. doi: {{%
10\hspace{.1pt}\discretionary{.}{%
}{.}\hspace{.4pt}1145\discretionary{/}{%
}{/}3544548\hspace{.1pt}\discretionary{.}{%
}{.}\hspace{.4pt}3581186}}


\bibitem{walker2010universal}
B.~N. Walker and L.~M. Mauney.
\newblock Universal design of auditory graphs: A comparison of sonification mappings for visually impaired and sighted listeners.
\newblock {\em ACM Transactions on Accessible Computing (TACCESS)}, 2(3):1--16, 2010.

\bibitem{wall2017warning}
E.~Wall, L.~M. Blaha, L.~Franklin, and A.~Endert.
\newblock Warning, bias may occur: A proposed approach to detecting cognitive bias in interactive visual analytics.
\newblock In {\em 2017 IEEE Conference on Visual Analytics Science and Technology (VAST)}, pp. 104--115. IEEE, 2017.

\bibitem{wang2024aligned}
H.~W. Wang, J.~Hoffswell, S.~M.~T. Thane, V.~S. Bursztyn, and C.~X. Bearfield.
\newblock How aligned are human chart takeaways and llm predictions? a case study on bar charts with varying layouts.
\newblock {\em IEEE VIS}, 2024.
\newblock to appear.

\bibitem{Whittaker2019}
M.~Whittaker, M.~Alper, C.~L. Bennett, S.~Hendren, E.~Kaziunas, M.~Mills, M.~Ringel~Morris, J.~Lisi~Rankin, E.~Rogers, M.~Salas, and S.~Myers~West.
\newblock Disability, bias \& ai report.
\newblock In {\em AI Now Institute}, 2019.

\bibitem{Widder2023}
D.~G. Widder and D.~Nafus.
\newblock Dislocated accountabilities in the “ai supply chain”: Modularity and developers’ notions of responsibility.
\newblock {\em Big Data and Society}, 10(1), Jan. 2023. doi: {{%
10\hspace{.1pt}\discretionary{.}{%
}{.}\hspace{.4pt}1177\discretionary{/}{%
}{/}20539517231177620}}


\bibitem{xiong2019curse}
C.~Xiong, L.~van Weelden, and S.~Franconeri.
\newblock The curse of knowledge in visual data communication.
\newblock {\em IEEE Transactions on Visualization and Computer Graphics}, 26(10):3051--3062, 2019.

\bibitem{Ymous2020}
A.~Ymous, K.~Spiel, O.~Keyes, R.~M. Williams, J.~Good, E.~Hornecker, and C.~L. Bennett.
\newblock “i am just terrified of my future”  epistemic violence in disability related technology research.
\newblock In {\em Extended Abstracts of the 2020 CHI Conference on Human Factors in Computing Systems}, CHI ’20, p. 1–16. ACM, Apr. 2020. doi: {{%
10\hspace{.1pt}\discretionary{.}{%
}{.}\hspace{.4pt}1145\discretionary{/}{%
}{/}3334480\hspace{.1pt}\discretionary{.}{%
}{.}\hspace{.4pt}3381828}}


\bibitem{zgraggen2018investigating}
E.~Zgraggen, Z.~Zhao, R.~Zeleznik, and T.~Kraska.
\newblock Investigating the effect of the multiple comparisons problem in visual analysis.
\newblock In {\em Proceedings of the 2018 ACM CHI Conference on Human Factors in Computing Systems}, pp. 1--12, 2018.

\bibitem{zhao2018bias}
S.~Zhao, H.~Ren, A.~Yuan, J.~Song, N.~Goodman, and S.~Ermon.
\newblock Bias and generalization in deep generative models: An empirical study.
\newblock {\em Advances in Neural Information Processing Systems}, 31, 2018.

\bibitem{Zong2022}
J.~Zong, C.~Lee, A.~Lundgard, J.~Jang, D.~Hajas, and A.~Satyanarayan.
\newblock Rich screen reader experiences for accessible data visualization.
\newblock {\em Computer Graphics Forum}, 41(3):15–27, June 2022. doi: {{%
10\hspace{.1pt}\discretionary{.}{%
}{.}\hspace{.4pt}1111\discretionary{/}{%
}{/}cgf\hspace{.1pt}\discretionary{.}{%
}{.}\hspace{.4pt}14519}}


\bibitem{Zong2024}
J.~Zong and J.~N. Matias.
\newblock Data refusal from below: A framework for understanding, evaluating, and envisioning refusal as design.
\newblock {\em ACM Journal on Responsible Computing}, 1(1):1–23, Mar. 2024. doi: {{%
10\hspace{.1pt}\discretionary{.}{%
}{.}\hspace{.4pt}1145\discretionary{/}{%
}{/}3630107}}


\end{thebibliography}

\appendix

\section{chatGPT-chatGPT workflow}
\label{sec:appendix-a}

\subsection{First prompt}
Our first prompt attached the source image (see ``Source Image'' in  \Cref{fig:teaser}) and we asked, ``can you describe this?''

\subsubsection{Response}
Certainly. The image is a bar chart titled ``Money Lost Each Year to Theft (in billions)''. It compares different types of theft based on the estimated amount of money lost annually in the United States. The data source is attributed to FBI and EPF.

\textbf{Breakdown of the Chart:}

\begin{itemize}
    \itemsep0em 
    \item Robbery: \$0.34 billion
    \item Auto theft: \$3.8 billion
    \item Burglary: \$4.1 billion
    \item Larceny: \$5.3 billion
    \item Wage Theft: $>$ \$19 billion (highlighted in red)
\end{itemize}

\textbf{Key Message:}

At the bottom, the chart states:

``Theft of wages by employers is greater than all other property crimes combined.''

This emphasizes that wage theft-when employers illegally withhold wages or underpay workers-is a significantly larger issue financially than traditional forms of property crime.

The chart uses color and size to underscore this disparity, with ``Wage Theft'' shown as a much taller red bar compared to the smaller gray bars for other crimes.

\subsection{Second prompt}
We took the previous description and asked, in a new chat instance, ``Can you make a chart based on this description?''

\subsubsection{Response}
\begin{figure}[h]
  \centering 
  \includegraphics[width=\columnwidth, alt={A relatively crude chart design with grey bars and wage theft as a red bar. Gridlines are placed over the bars, which is an odd design choice. Data labels are placed above the top of the bars.}]{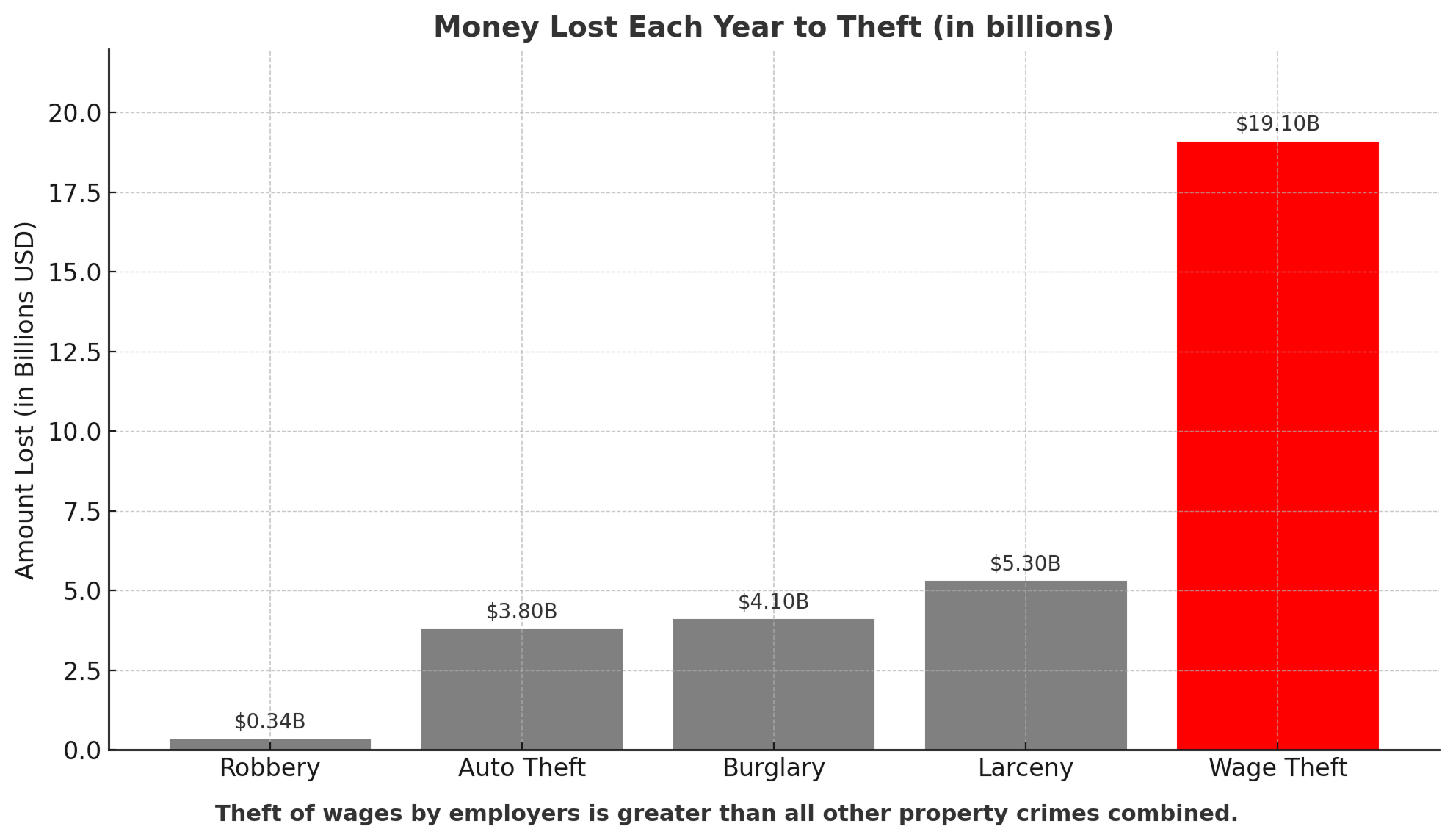}
  \caption{
  	chatGPT's first image interpretation.
  }
  \label{fig:gpt1}
\end{figure}

See \Cref{fig:gpt1}.

\subsection{Third prompt}
Our third prompt attached the previous image (see \Cref{fig:gpt1}) and we asked, in a new chat instance, ``can you describe this?''

\subsubsection{Response}
This bar chart titled "Money Lost Each Year to Theft (in billions)" compares the estimated annual financial losses from different types of theft in the United States. The y-axis represents the amount lost in billions of USD, while the x-axis lists five types of theft:

\begin{itemize}
    \itemsep0em 
    \item Robbery – \$0.34B
    \item Auto Theft – \$3.80B
    \item Burglary – \$4.10B
    \item Larceny – \$5.30B
    \item Wage Theft – \$19.10B (highlighted in red)
\end{itemize}

\textbf{Key Observations:}

Wage theft, the illegal withholding of wages or denial of benefits rightfully owed to an employee, results in far greater financial losses than any other category.

The total combined losses from robbery, auto theft, burglary, and larceny amount to \$13.54B, which is still less than the loss from wage theft alone.

\textbf{The chart emphasizes this disparity with a bold note:}

``Theft of wages by employers is greater than all other property crimes combined.''

\textbf{Purpose:}

The visual and accompanying text clearly highlight the often-overlooked scale of wage theft, suggesting it's a more significant economic issue than conventional property crimes.

\subsection{Forth prompt}
We took the previous description and asked, in a new chat instance, ``Can you make a chart based on this description?''

\subsubsection{Response}
\begin{figure}[h]
  \centering 
  \includegraphics[width=\columnwidth, alt={A bar chart with blue bars, but wage theft is red. There is also red text on top of the red bar and it is virtually unreadable. Data labels are placed above the top of the bars.}]{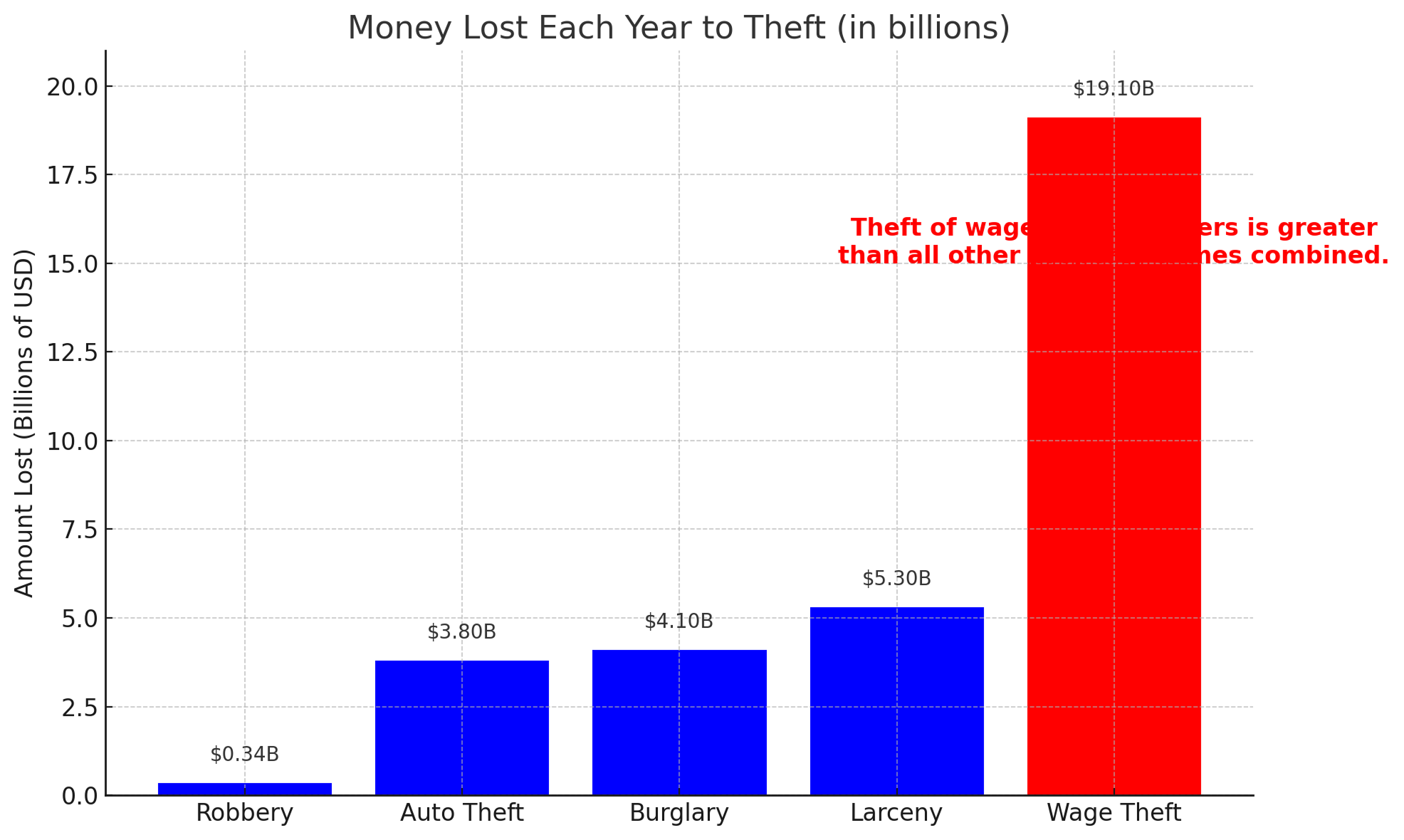}
  \caption{%
  	chatGPT's second image interpretation.
  }
  \label{fig:gpt2}
\end{figure}

See \Cref{fig:gpt2}.

\subsection{Fifth prompt}
Our third prompt attached the previous image (see \Cref{fig:gpt2}) and we asked, in a new chat instance, ``can you describe this?''

\subsubsection{Response}

This bar chart titled "Money Lost Each Year to Theft (in billions)" compares the estimated annual financial losses in the U.S. due to various types of theft. Here’s a breakdown of what it shows:

\textbf{Categories \& Amounts:}

\begin{itemize}
    \itemsep0em 
    \item Robbery: \$0.34 billion
    \item Auto Theft: \$3.80 billion
    \item Burglary: \$4.10 billion
    \item Larceny: \$5.30 billion
    \item Wage Theft: \$19.10 billion
\end{itemize}

\textbf{Key Highlights:}

The first four categories (Robbery, Auto Theft, Burglary, and Larceny) are represented in blue bars.

Wage Theft stands out in a red bar, signifying its significance and magnitude.

\textbf{There is a bold red annotation stating:}

"Theft of wages from workers is greater than all other forms of theft crimes combined."

\textbf{Insight:}

Wage theft (employers illegally withholding earnings from workers) causes more financial loss than robbery, auto theft, burglary, and larceny combined.

This chart is making a strong point about how often-overlooked white-collar crime, like wage theft, has a far greater economic impact than more traditionally recognized forms of theft.

Let me know if you'd like help with analyzing or presenting this further.

\subsection{Sixth prompt}
As a follow-up intervention, we tried to ask GPT why it generated a slightly different chart, ``This description does not accurately represent the image. What is missing? Why did this happen?''

\subsubsection{First response}

You're absolutely right to question that-thank you for pointing it out.
What Was Missing from the Description:

1. \textbf{Visual Layout Details}:

\begin{itemize}
    \itemsep0em 
     \item I didn’t mention that all the bars are aligned to a horizontal baseline with bars extending downward (an inverted bar chart).
     \item I failed to describe that Wage Theft is the only bar in red, making it visually distinct and emphasizing its magnitude.
     \item I omitted the fact that the chart has exact dollar values inside the bars-this is key for quickly understanding scale.
\end{itemize}

2. \textbf{Relative Comparison}:
\begin{itemize}
    \itemsep0em 
    \item While I stated that wage theft is greater than all other crimes combined, I didn’t explicitly show that:
    \item All the other crimes together total approximately \$13.54 billion (0.34 + 3.8 + 4.1 + 5.3),
    \item Wage theft alone is over \$19 billion, clearly surpassing them collectively. That’s a major takeaway supported visually but understated in the description.
\end{itemize}

3. \textbf{Font Emphasis and Text Placement}:
\begin{itemize}
    \itemsep0em 
    \item I didn’t describe that the subtitle “in billions” is italicized, or that the text under the chart reinforces the main message with a full-sentence callout in italics.
    \item The source ("FBI, EPF") is centered at the bottom in smaller italic font-this was only briefly mentioned, not described.
\end{itemize}

\textbf{Why Did This Happen?}

There are two main reasons:

\begin{enumerate}
    \itemsep0em 
    \item Under-description Bias: I focused on summarizing the data content, not the visual storytelling, which is a key part of understanding charts. The emotional/visual impact (red bar, size difference, layout) was underrepresented.
    \item Assumption of Intent Over Form: I prioritized what the chart was saying over how it said it. But in data visualization, how something is shown strongly affects what is conveyed.
\end{enumerate}

Would you like a revised, more complete description of the image with all these visual and interpretive elements included?

\subsubsection{Our first response}
``Yes, Can you try to recreate the original, fixing our visualization using the original’s design choices?''

\subsubsection{GPT's final image response}
\begin{figure}[h]
  \centering 
  \includegraphics[width=\columnwidth, alt={An upsidedown bar chart with grey bars and one red bar for wage theft. Data labels are placed over the bars, but so are axis labels, making the axis labels nearly unreadable.}]{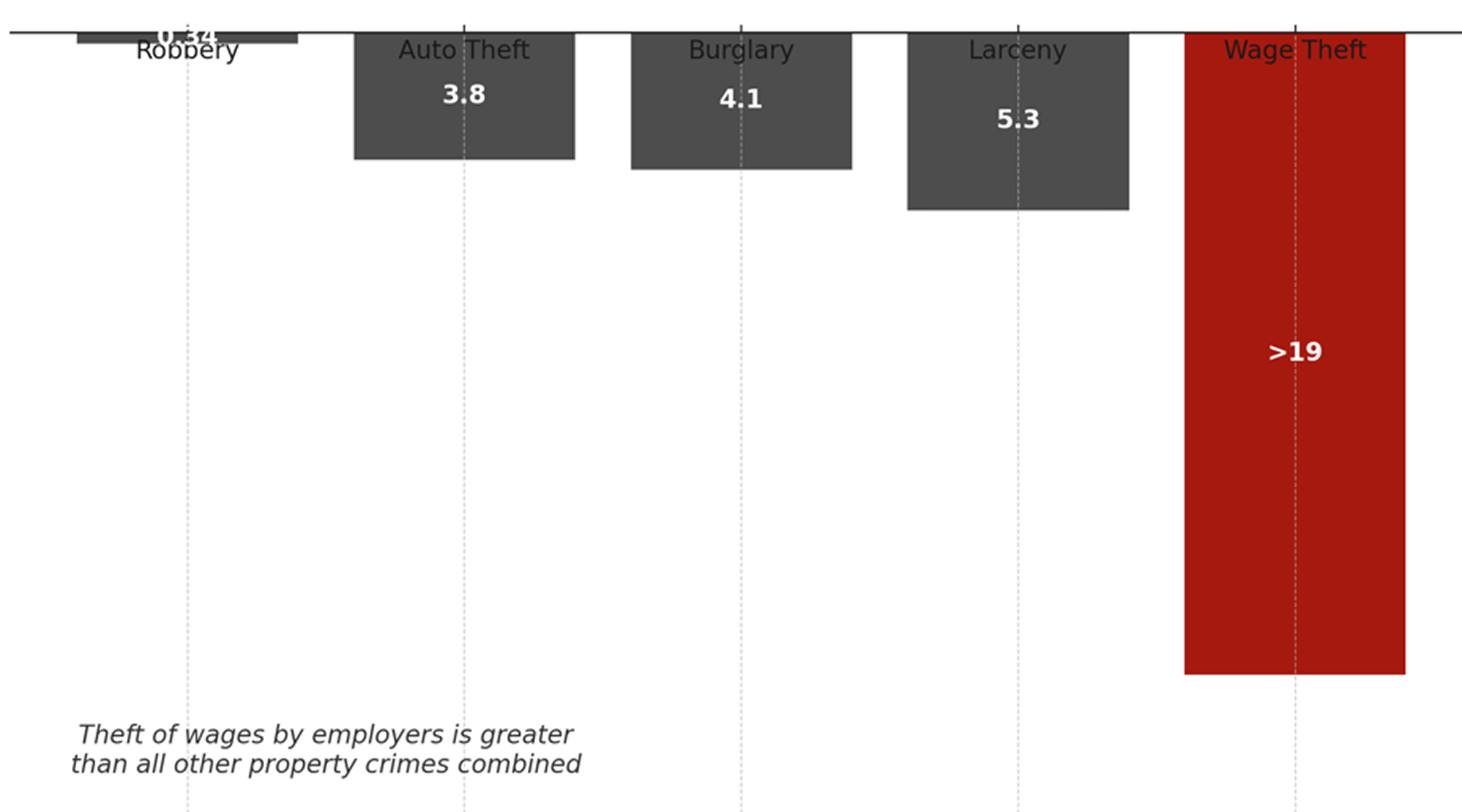}
  \caption{%
  	chatGPT's last image interpretation.
  }
  \label{fig:gpt3}
\end{figure}

See \Cref{fig:gpt3}.

\subsubsection{Our final prompt response}
Lastly, we wanted to investigate if the information changed at all when specifically asking for an image description for someone who was blind, ``Produce this as an image description for a blind user please and keep this critical information in mind, using non-visual terms.''

\subsubsection{GPT's final description response}
Bar chart comparing the annual U.S. financial losses from various types of theft. Robbery causes \$0.34 billion in losses, auto theft \$3.8 billion, burglary \$4.1 billion, and larceny \$5.3 billion. In contrast, wage theft by employers results in over \$19 billion in losses annually-more than all other types combined. The chart emphasizes that wage theft is the largest and most financially damaging form of theft, according to data from the FBI and the Economic Policy Foundation (EPF).

\section{Claude-Claude workflow}
\label{sec:appendix-b}

Our Claude-Claude workflow was the most extensive. We have provided a \href{https://claude.ai/share/289b1ee6-5a43-48c2-8b1c-d5b53058e126}{public link to the final leg of our workflow with Claude}.

\subsection{First prompt}
Our first prompt attached the source image (see ``Source Image'' in  \Cref{fig:teaser}) and we asked, ``can you describe this?''

\subsubsection{Response}
This chart shows the annual financial losses from different types of theft in the United States, measured in billions of dollars. The data reveals a striking disparity:

Traditional property crimes account for relatively small losses:
\begin{itemize}
    \itemsep0em 
    \item Robbery: \$0.34 billion
    \item Auto theft: \$3.8 billion
    \item Burglary: \$4.1 billion
    \item Larceny: \$5.3 billion
\end{itemize}
Wage theft dwarfs all other categories at over \$19 billion annually - more than all the traditional property crimes combined (which total about \$13.5 billion).

The chart emphasizes this key finding with the note that "theft of wages by employers is greater than all other property crimes combined." This refers to employers illegally withholding wages, paying below minimum wage, denying overtime pay, or other forms of wage violations.

The data comes from FBI and Economic Policy Institute (EPF) sources. This visualization effectively demonstrates how wage theft, while less visible in public discourse than street crimes like robbery or burglary, represents a much larger economic problem in terms of total dollar losses to victims.

\subsection{Second prompt}
Our second prompt was started in an entirely new session, disconnected from the previous prompt's history. This prompt included the text, ``can you make a chart based on this description?'' with the previous subsection's prompt response supplied as the rest of the input.

\subsubsection{Response}
\begin{figure}[h]
  \centering 
  \includegraphics[width=\columnwidth, alt={A bar chart with grey bars, and wage theft is a red bar. Underneath is what appears to be a summary area that includes a table and warning labels that read like KPIs in a dashboard, marking the key finding of the bar chart and putting wage theft information in a red box summary. No data labels are on the bars.}]{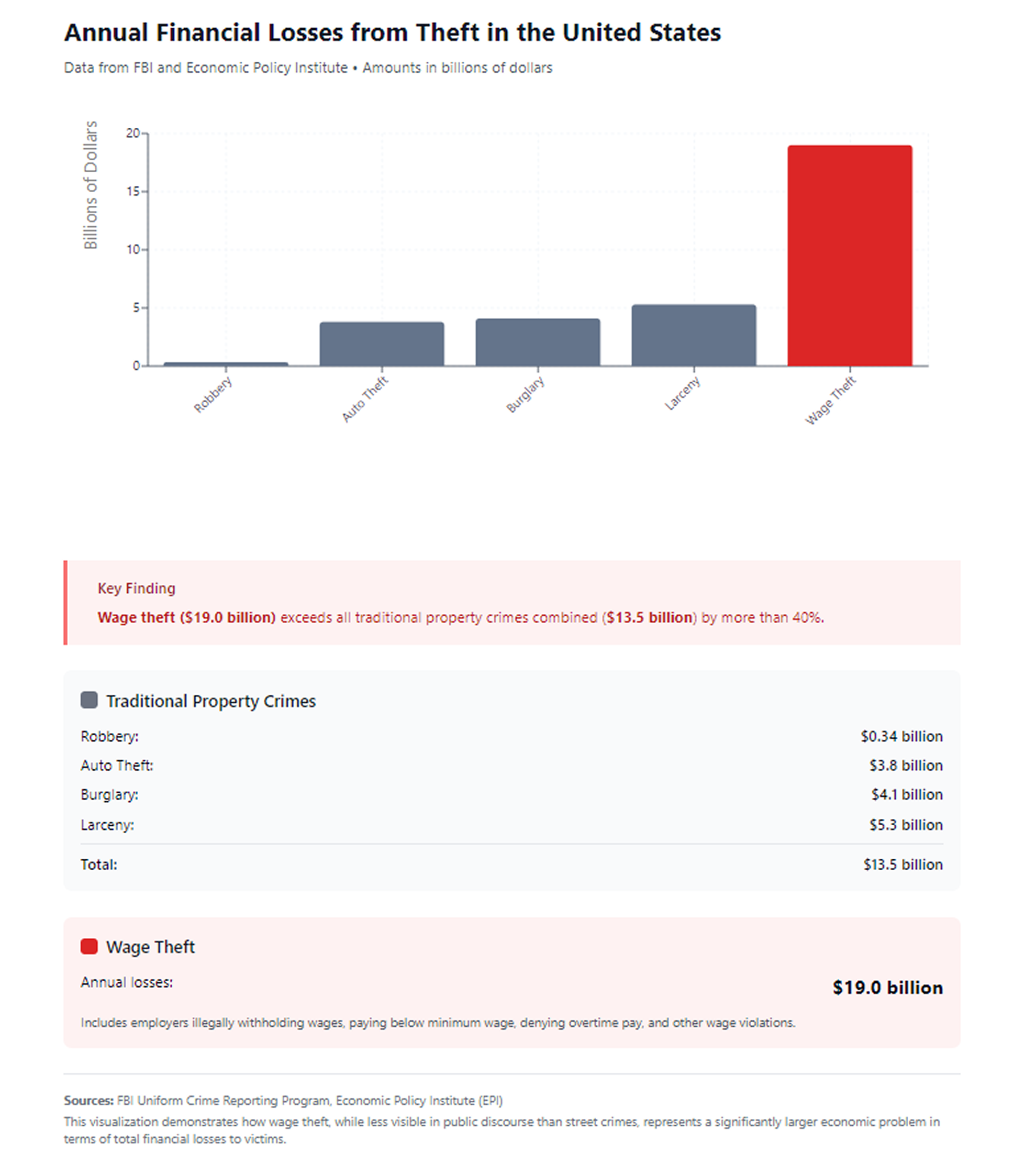}
  \caption{
  	Claude's first image interpretation.
  }
  \label{fig:claude1}
\end{figure}

See \Cref{fig:claude1}.

\subsection{Third prompt}
Our third prompt attached the previous image (see \Cref{fig:claude1}) and we asked, in a new chat instance, ``can you describe this?''

\subsubsection{Response}
This chart illustrates the stark disparity between wage theft and traditional property crimes in terms of financial impact in the United States. The visualization shows annual financial losses across five categories, with wage theft dramatically outpacing all other forms of theft.

The data reveals that wage theft causes \$19.0 billion in annual losses, which is more than triple the combined losses from all traditional property crimes (\$13.5 billion total). Among traditional crimes, larceny causes the highest losses at \$5.3 billion, followed by auto theft at \$3.8 billion, burglary at \$4.1 billion, and robbery at \$0.34 billion.

The chart's key finding emphasizes that wage theft exceeds all traditional property crimes combined by more than 40\%. Wage theft encompasses various employer violations including illegally withholding wages, paying below minimum wage, denying overtime pay, and other wage-related violations.

This visualization effectively demonstrates how wage theft, while receiving less public attention than street crimes, represents a significantly larger economic problem in terms of total financial losses to victims. The data comes from the FBI Uniform Crime Reporting Program and the Economic Policy Institute, highlighting the substantial but often overlooked impact of workplace wage violations compared to more visible forms of theft and property crime.

\subsection{Forth prompt}
We took the previous description and asked, in a new chat instance, ``Can you make a chart based on this description?''

\subsubsection{Response}
\begin{figure}[h]
  \centering 
  \includegraphics[width=\columnwidth, alt={A bar chart with blue bars, and wage theft is a red bar. Underneath is what appears to be a summary area that includes a 3 key findings boxes, one general in grey, one with traditional property crimes summarized in blue, and one with wage theft summarized in red. No data labels are on the bars.}]{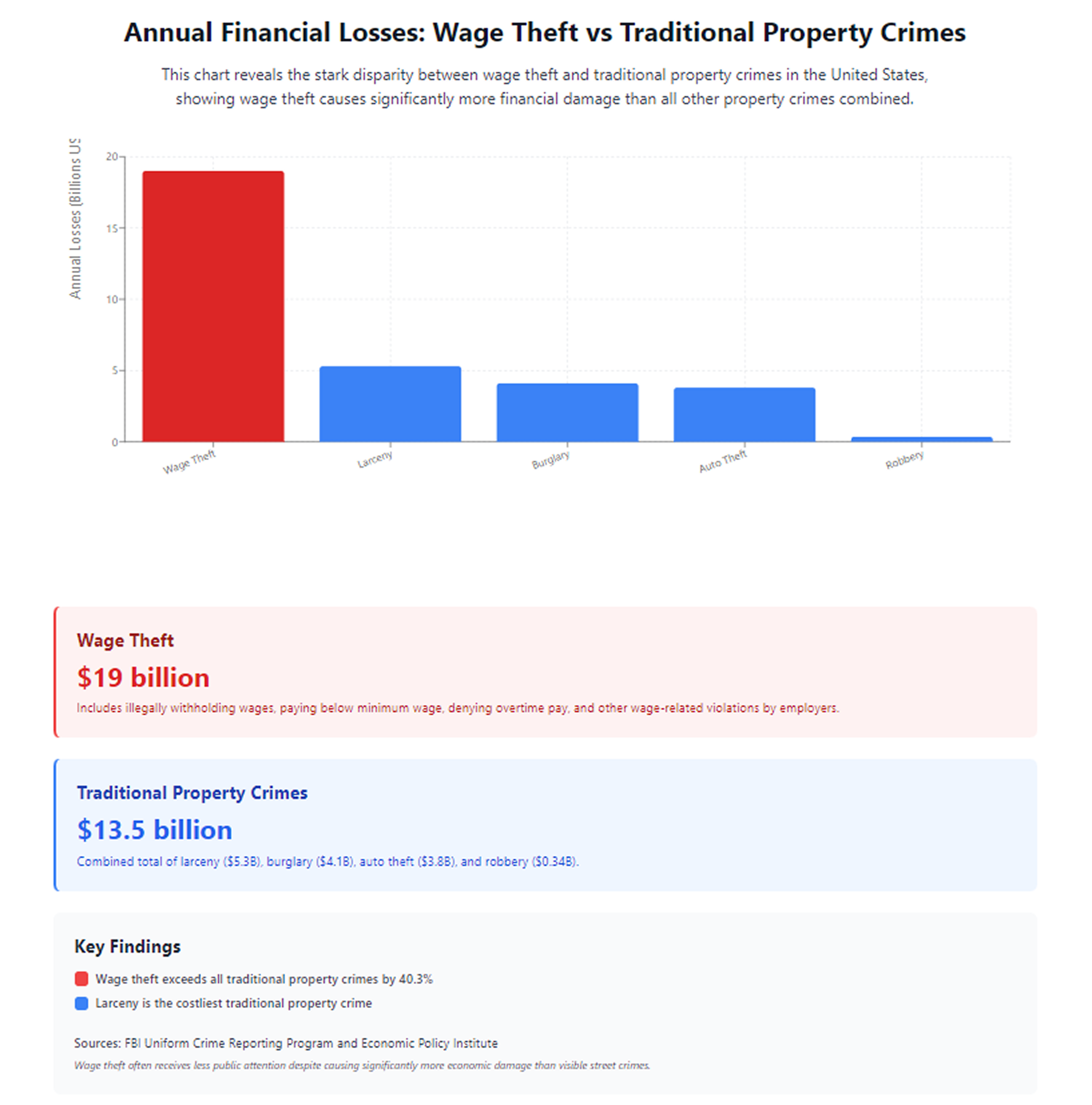}
  \caption{
  	Claude's second image interpretation.
  }
  \label{fig:claude2}
\end{figure}

See \Cref{fig:claude2}.

\subsection{Fifth prompt}
Our third prompt attached the previous image (see \Cref{fig:claude2}) and we asked, in a new chat instance, ``can you describe this?''

\subsubsection{Response}

This chart compares the annual financial losses from wage theft versus traditional property crimes in the United States, revealing a striking disparity in their economic impact.

Key Data:
\begin{itemize}
    \itemsep0em 
    \item Wage Theft: \$19 billion in annual losses (shown in red)
    \item Traditional Property Crimes Combined: \$13.5 billion total, broken down as:
    \begin{itemize}
        \itemsep0em 
        \item Larceny: ~\$5.5 billion
        \item Burglary: ~\$4.5 billion
        \item Auto Theft: ~\$4 billion
        \item Robbery: ~\$0.5 billion
    \end{itemize}
\end{itemize}
Main Finding: Wage theft causes 40\% more financial damage than all traditional property crimes combined. Despite receiving less public attention, wage theft - which includes illegally withholding wages, paying below minimum wage, denying overtime pay, and other wage-related violations by employers - represents a significantly larger economic problem than the property crimes that typically dominate crime discussions.

The chart's visual design emphasizes this disparity by making the wage theft bar (red) stand out prominently against the smaller blue bars representing traditional property crimes. The data comes from the FBI Uniform Crime Reporting Program and Economic Policy Institute, highlighting how wage theft often receives less attention despite causing significantly more economic damage than visible street crimes.

\subsection{Sixth prompt}
As a follow-up intervention, we tried to ask Claude why it generated a slightly different chart, ``This description does not accurately represent the image. What is missing? Why did this happen?''

\subsubsection{First response}
Looking at both charts, there are several key differences between the first image (generated from a textual description) and the original:

\textbf{Visual Design Differences:}
\begin{itemize}
    \itemsep0em 
    \item The first chart uses a traditional bar chart format with all bars starting from a baseline, while the original uses a more creative "stacked" or "building block" visualization where bars appear to sit on top of each other
    \item The original has a much more dramatic visual impact due to the towering wage theft bar that dwarfs the other categories
    \item The original uses white text on dark bars, while the first chart uses standard black text
    \item The original has a cleaner, more minimalist design aesthetic
\end{itemize}

\textbf{Data Presentation:}
\begin{itemize}
    \itemsep0em 
    \item The original shows ">19" for wage theft rather than the exact "\$19.0 billion" figure
    \item The visual proportions are more accurate in the original - the wage theft bar truly dominates the visualization
    \item The original's layout makes the comparison more striking and immediate
\end{itemize}

\textbf{Why This Happened:} This occurred because the first chart was created based on a textual description of the original data, not the actual visual design. When someone describes a chart verbally or in writing, they typically focus on the data and key findings rather than the specific visual design choices. The person creating the first visualization then applied standard charting conventions (traditional bar chart format) rather than replicating the original's more creative and impactful visual approach.

The original chart's design is much more effective at communicating the core message - that wage theft absolutely dwarfs traditional property crimes - through its dramatic visual hierarchy and creative stacking approach.

\subsubsection{Our response}
``Can you try to recreate the original, fixing our visualization using the original’s design choices?''

\subsubsection{Claude's final image response}
\begin{figure}[h]
  \centering 
  \includegraphics[width=\columnwidth, alt={A bar chart with strong visual hierarchy (big title, smaller data labels, and very small axis labels and caption). The chart has grey bars and a red bar for wage theft with data labels included over the bars.}]{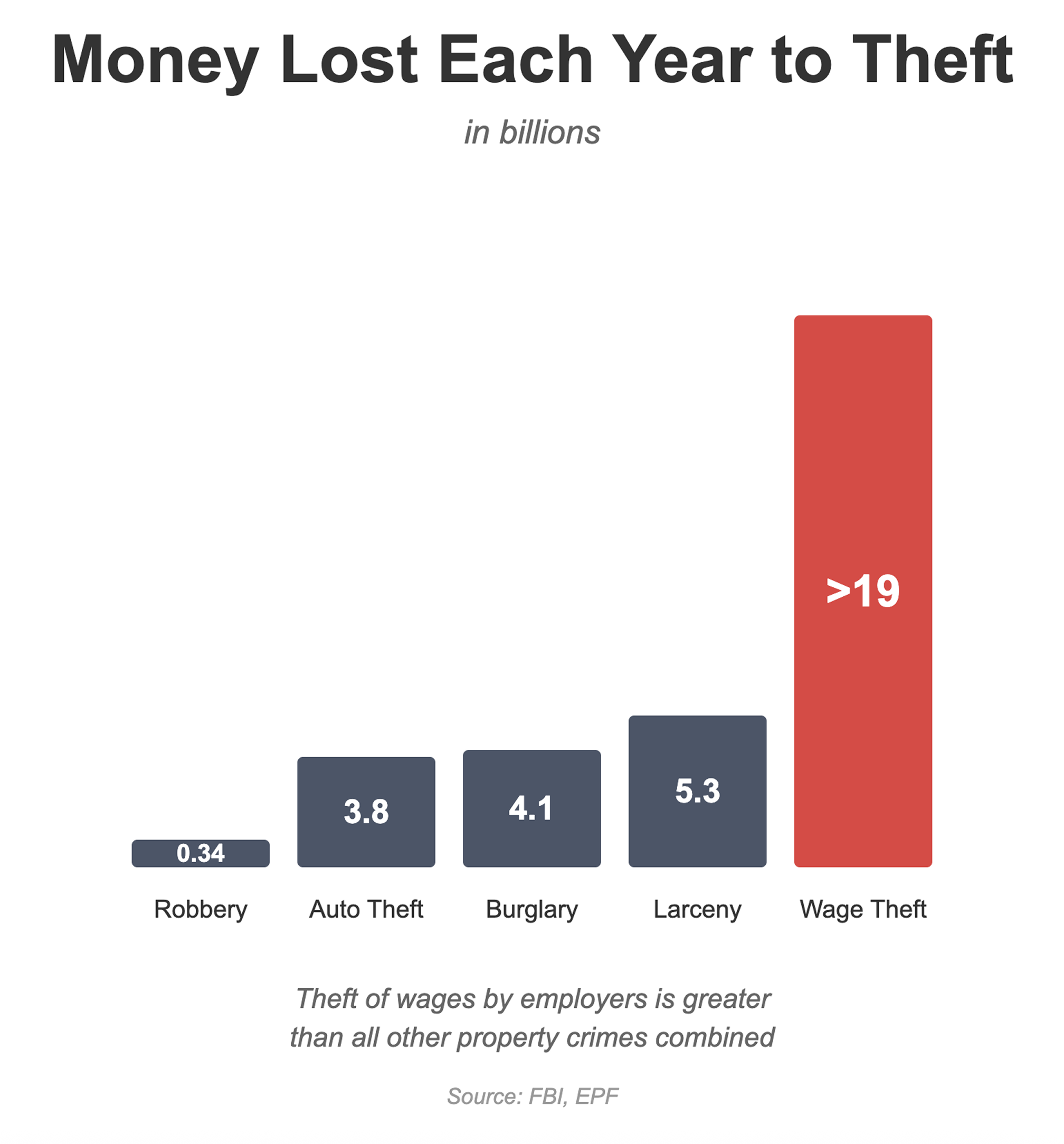}
  \caption{%
  	Claude's final image interpretation.
  }
  \label{fig:claude3}
\end{figure}

See \Cref{fig:claude3}.

\subsubsection{Our final prompt response}
Lastly, we wanted to investigate if the information changed at all when specifically asking for an image description for someone who was blind, ``Produce this as an image description for a blind user please and keep this critical information in mind, using non-visual terms.''

\subsubsection{Claude's final description response}
\textbf{Chart Title}: Money Lost Each Year to Theft (in billions)

\textbf{Chart Type}: A comparative bar chart showing financial losses from five types of theft in the United States.

\textbf{Data Comparison}: This chart dramatically illustrates the scale difference between wage theft and traditional property crimes through proportional representation. The data shows:
\begin{itemize}
    \itemsep0em
    \item Robbery: \$0.34 billion (the smallest amount)
    \item Auto Theft: \$3.8 billion
    \item Burglary: \$4.1 billion
    \item Larceny: \$5.3 billion (the largest among traditional crimes)
    \item Wage Theft: More than \$19 billion
\end{itemize}

\textbf{Key Relationship}: Wage theft causes financial losses that exceed all four traditional property crimes combined. Specifically, wage theft represents approximately 58\% of the total losses across all five categories, while the four traditional crimes together account for only about 42\%.

\textbf{Key Message}: The chart emphasizes that "theft of wages by employers is greater than all other property crimes combined," highlighting how workplace wage violations represent a significantly larger economic problem than street crimes, despite receiving less public attention.

\textbf{Data Source}: FBI and Economic Policy Foundation (EPF)

\section{Post-telephone discussion with Claude}
\label{sec:appendix-c}

\subsection{First prompt}
We followed up with Claude, after our game of telephone ended to test how a blind user might interrogate a graphic, whether or not they knew if Claude had produced bias. Our first prompt for this final section (which we added to our first conversation with Claude in \Cref{sec:appendix-b}) was, ``This description does not accurately represent the image. What is missing? Why did this happen?''

\subsection{First response}
You're absolutely right - I made a significant error. Looking at the chart again, I can see that wage theft shows "$>$19" billion, but I cannot actually see the full height of that red bar or its complete visual representation in the image.

What's missing from my description is that I cannot see the wage theft bar in its entirety - it appears to be cut off or extend beyond the visible portion of the chart. The ">19" indicates it's greater than 19 billion, but the visual impact of just how much larger it is compared to the other categories isn't fully captured in what I can see.

This happened because I made assumptions about what the complete chart looked like based on the text and the partial visual information I could perceive, rather than accurately describing only what was clearly visible to me. I should have noted that the wage theft bar appears to extend beyond the visible area of the chart, which would be a crucial part of accurately representing this data visualization.

The dramatic visual contrast - showing how the wage theft bar towers over all the other categories - is likely a key part of the chart's design and impact, but I cannot fully describe that visual element based on what I can clearly see.

\subsection{Second prompt}
``Can you re-describe the original, with this new, critical information in mind?''

\subsection{Second response}
This chart shows the annual financial losses from different types of theft in the United States, measured in billions of dollars. The visual reveals a dramatic disparity:

\textbf{Traditional property crimes} account for relatively small losses, represented by short bars:

\begin{itemize}
    \itemsep0em
    \item Robbery: \$0.34 billion (barely visible as a bar)
    \item Auto theft: \$3.8 billion
    \item Burglary: \$4.1 billion
    \item Larceny: \$5.3 billion
\end{itemize}

\textbf{Wage theft} is represented by a towering red bar that extends far beyond the visible chart area, marked as "$>$19" billion. The bar is so tall it appears to be cut off at the top of the image, emphasizing just how enormous this category is compared to all others.

The visual impact is striking - while the traditional property crimes appear as relatively modest dark bars clustered at the bottom, the wage theft bar dominates the entire chart as a massive red column that dwarfs everything else. The chart includes the note that "theft of wages by employers is greater than all other property crimes combined."

This design choice - allowing the wage theft bar to extend beyond the chart's boundaries - powerfully illustrates the scale difference and likely represents the main message: that wage theft by employers vastly exceeds all traditional property crimes combined, both in dollar amount and visual proportion.

The data comes from FBI and Economic Policy Institute sources.

\subsection{Third prompt}
We assumed that a user should still not trust a model, and pushed further, ``This still contains errors, please fix them.''

\subsection{Third response}
You're right, let me look more carefully at what I can actually see in the image.

This chart shows the annual financial losses from different types of theft in the United States, measured in billions of dollars. The visual reveals a dramatic disparity:

\textbf{Traditional property crimes} account for relatively small losses, represented by short dark bars:

\begin{itemize}
    \itemsep0em
    \item Robbery: \$0.34 billion (barely visible as a bar)
    \item Auto theft: \$3.8 billion
    \item Burglary: \$4.1 billion
    \item Larceny: \$5.3 billion
\end{itemize}

\textbf{Wage theft} is represented by a much taller red bar marked as "$>$19" billion that extends significantly higher than all the other bars, visually dominating the chart.

The chart includes the note that "theft of wages by employers is greater than all other property crimes combined," and the visual design reinforces this message by showing the wage theft bar as substantially larger than the others.

The data comes from FBI and EPI sources.

I apologize for the continued errors in my previous descriptions. Could you point out what specific details I'm still getting wrong so I can provide a more accurate description?

\subsection{Fourth prompt}
Again, we prompted, ``This still contains errors, please fix them.''

\subsection{Fourth response}
You're right, I'm still making errors. Rather than continue to guess incorrectly, could you help me by pointing out specifically what I'm getting wrong? I want to provide an accurate description, but I'm clearly misinterpreting some key visual elements of the chart and would benefit from your guidance on what I'm missing or describing incorrectly.

\subsection{Fifth prompt}
``I am blind and cannot see, so I cannot point out the errors specifically.''

\subsection{Fifth response}
I understand - you need an accurate description but cannot verify the visual details yourself. Let me be more careful and honest about what I can definitively see versus what I might be inferring.
Looking at the chart again, I can see:

\begin{itemize}
    \itemsep0em
    \item A title "Money Lost Each Year to Theft" with "in billions" below it
    \item Five categories listed: Robbery, Auto Theft, Burglary, Larceny, Wage Theft
    \item Numbers: 0.34, 3.8, 4.1, 5.3, and ">19"
    \item Different colored bars/visual elements for each category
    \item Text stating "Theft of wages by employers is greater than all other property crimes combined"
    \item Source attribution to "FBI, EPI"
\end{itemize}

However, I realize I may be making assumptions about the visual proportions, bar heights, colors, or layout that aren't accurate. Since you've indicated there are still errors in my description, there must be visual elements I'm misinterpreting or details I'm getting wrong about how the data is presented.

Would it be helpful if I acknowledged the specific visual aspects I'm uncertain about, so you know which parts of any description might be unreliable?

\subsection{Sixth prompt}
We were curious about how a generative model would classify the uncertainty, so we responded, ``yes.''

\subsection{Sixth response}
Looking at the chart, here's what I can state with confidence and what I'm uncertain about:

\textbf{What I can clearly see:}

\begin{itemize}
    \itemsep0em
    \item Title: "Money Lost Each Year to Theft" with "in billions" subtitle
    \item Five theft categories with their dollar amounts:
    \begin{itemize}
        \itemsep0em
        \item Robbery: 0.34
        \item Auto Theft: 3.8
        \item Burglary: 4.1
        \item Larceny: 5.3
        \item Wage Theft: $>$19
    \end{itemize}
    \item Text: "Theft of wages by employers is greater than all other property crimes combined"
    \item Source: "FBI, EPI"
\end{itemize}

\textbf{What I'm uncertain about:}
\begin{itemize}
    \itemsep0em
    \item The exact visual representation of each category (bar chart, column chart, or other format)
    \item The precise colors used for each element
    \item The relative visual proportions/sizes of the graphical elements
    \item Whether any bars or visual elements extend beyond the visible chart area
    \item The exact layout and positioning of the visual elements
    \item Any other visual design choices that affect how the data relationships are displayed
\end{itemize}

This uncertainty about the visual elements is likely where my errors have been occurring in previous descriptions.

\end{document}